\def\be{\begin{equation}}
\def\ee{\end{equation}}
\def\bea{\begin{eqnarray}}
\def\eea{\end{eqnarray}}
\def\a{\alpha}
\def\b{\beta}
\def\d{\delta}
\def\e{\epsilon}
\def\L{${\cal L}$ }
\def\t{\phi}
\def\tn{\phi_n}
\def\p{\theta}
\def\tm{\theta_m}
\def\x{\psi}
\def\l{\lambda}
\documentclass[12pt]{article}

\begin{document}
\title{Theory of  resistor networks: The two-point resistance}
\author{ F. Y. Wu\\
Department of Physics,\\ 
Northeastern University \\
Boston, Massachusetts, 02115 }
\date{}
\maketitle
\begin{abstract}
The resistance between arbitrary two nodes in a resistor network
is obtained in terms of the eigenvalues and eigenfunctions of the Laplacian matrix
associated with the network.  Explicit formulas  for two-point resistances are deduced
for regular lattices  in one, two, and three dimensions under various boundary conditions
including that of a M\"obius strip and a Klein bottle.  The emphasis is on lattices of finite sizes.
We also deduce  summation and product identities which can be used to analyze large-size
expansions  of two-and-higher dimensional lattices.
\end{abstract}
 
\newpage
\section{Introduction}
A classic problem in electric circuit theory 
studied by numerous authors over many years  is the
 computation of the resistance between two nodes in a resistor network
(for a list of relevant
references up to 2000 see, e.g., \cite{cserti}).  
Besides being a central problem in  electric circuit theory,
the computation of resistances is also relevant to 
a wide range of problems ranging from random walks (see \cite{ds,lovasz} and discussions below), 
the theory of harmonic functions \cite{pol}, first-passage processes \cite{redner},
to lattice Green's functions \cite{lg}.
The connection with these problems originates from the fact that  
electrical potentials on a grid are governed by the same difference
equations as those occurring in the other problems. 
For this reason, the 
 resistance problem 
is often studied from the point of view of solving the
difference equations, which is most conveniently carried out for
 infinite networks.  
Very little attention has been paid to finite 
networks, even though the latter are the ones  occurring in real life.
In this paper we take up this problem and present
a general formulation for
finite networks. Particularly, known results for infinite networks are recovered
by taking the infinite-size limit.  In later papers we plan to study effects
of lattice defects and carry out finite-size analyses.
 
\medskip
The study of electric networks was formulated by
  Kirchhoff \cite{kirch} more than 150 years ago
as an instance of a linear analysis. 
 Here, we analyze the  problem along the same line of approach.
Our starting point is the consideration of the Laplacian matrix associated with
a network, which is a matrix  whose entries are 
the conductances connecting pairs of nodes.
 Just as in graph theory that everything about a graph is 
described by  its adjacency matrix (whose elements is 1 if two vertices are
connected and 0 otherwise), we expect everything about an electric network
is described by its  Laplacian.
   Indeed, in Section 2 below we shall derive
an expression of  the two-point resistance between arbitrary two nodes in terms of 
 the eigenvalues and eigenvectors of the Laplacian matrix \cite{note}.  
 In ensuing sections we apply 
our formulation to  networks of  one-dimensional and 
two-dimensional lattices under various  boundary conditions
 including those embedded on a
 M\"obius strip and a Klein bottle, and lattices in higher spatial dimensions. 
We also deduce   summation and product identities which can be used
to reduce the computational labor as well as analyze large-size expansions  
in two-and-higher dimensions.
  In subsequent papers  we shall consider
 large-size expansions and effects of defects in finite networks, the latter a
 problem that
 has been studied in the past only for infinite networks \cite{cserti1}.

\medskip
 Let  \L be a resistor network consisting of ${\cal N}$ nodes numbered  by $i=1,2,\cdots,{\cal N}$. 
Let $r_{ij}=r_{ji}$ be the resistance of the
 resistor connecting   nodes $i$ and $j$.  Hence,
  the  conductance is
\bea
c_{ij} = r_{ij}^{-1} = c_{ji}  \nonumber
\eea
so that
 $c_{ij}=0$ (as in an adjacency matrix)
if there is no resistor connecting $i$ and $j$.

\medskip
Denote the electric potential at the $i$th node
by $V_i$  and  the {net} current flowing {\it into}  the network at the $i$th node by
$I_i$ (which is zero if the $i$th node is
not connected to the external world).  Since there
exist no sinks or sources of current including the external world, we have the constraint
\be
\sum_{i=1}^{\cal N} I_i=0. \label{sumrule}
\ee

The Kirchhoff law states 
\be
\sum^{\cal N}_{j=1} \ '\ c_{ij} \big( V_i - V_j \big) = I_i,
\hskip 1cm i=1,2,\cdots, {\cal N}, \label{eq1}
\ee
where the prime denotes the omission of the term $j=i$.
Explicitly, (\ref{eq1}) reads
\be
{\bf L}\ {\vec V} = {\vec I} \label{eq3}
\ee
where
\bea
{\bf L} &=& 
\pmatrix {c_1&-c_{12}& \dots &-c_{1{\cal N}}\cr
                    -c_{12} & c_2 & \dots & -c_{2{\cal N}} \cr
                     \vdots & \vdots & \ddots & \vdots & \cr
                     -c_{1{\cal N}} & -c_{2{\cal N}} & \dots & c_{{\cal N}} \cr} \label{eq4}
\eea
and
\bea
c_i \equiv \sum^{\cal N}_{j=1} \ '\ c_{ij}. \label{Xdef}
\eea
Here ${\vec V}$ and ${\vec I}$ are ${\cal N}$-vectors whose components are $V_i$ and $I_i$ respectively.
We note in passing that 
if all nonzero resistances are equal to 1, then the off-diagonal
elements of the matrix $-{\bf L}$  are precisely 
those of  the adjacency matrix of ${\cal L}$.

\medskip
 The Laplacian matrix ${\bf L}$ 
of the network ${\cal L}$ is also known as the Kirchhoff  matrix,
or simply the tree matrix; the latter name derived
from the fact  that  all cofactors of  ${\bf L}$ are equal, 
and equal to the
spanning tree generating function for ${\cal L}$ \cite{fara}.

\medskip
To compute  the resistance
$R_{\a\b}$ between two  nodes $\a$ and $\b$, we connect $\a$ and $\b$
to the two terminals of an external battery and measure the current $I$ going through
the battery while no other nodes are connected to external sources. 
Let the potentials at the two nodes be, respectively,
$V_\a$ and $V_\b$.   Then, the desired resistance is the ratio
\be
R_{\a\b} = {{V_\a-V_\b}\over I}.\label{equalr}
\ee
The computation of the two-point resistance $R_{\a\b}$ 
is now reduced to solving (\ref{eq3}) for $V_\a$ and $V_\b$
 with the  current given by
 \bea
I_i =I(\d_{i\a} -\d_{i\b}) . \label{I}
\eea
   
\noindent
{\bf A probabilistic interpretation}: 

\medskip
The two-point resistance has a 
probabilistic interpretation.  Consider a random walker walking
on the network ${\cal L}$ with the probability 
\be
p_{i\to j} = c_{ij}/c_i \label{prob}
\ee
of hoping
 from node $i$ to node $j$, where $p_{i\to j}$ can be different from  $p_{j\to i}$.
Let $P(\a ;\b)$ be the probability that the walker starting from node $\a$
will reach node $\b$ before returning to $\a$, 
which is the probability of first passage.  Then one has
the relation  \cite{ds,redner}
\bea
P(\a;\b)= \frac 1 {c_\a R_{\a\b}}  \label{probab}
    \eea
where $c_\a$ is defined in (\ref{Xdef}).  If all resistances are 1, then
(\ref{probab}) becomes
\bea
P(\a;\b) = \frac 1 {z_\a R_{\a\b}}.
  \label{probab1}
\eea
where $z_\a$ is the 
coordination number of, or the number of nodes connected to, the node $\a$.

\section{The two-point resistance: A theorem}
Let $\Psi_i$ and $\l_i$ be the eigenvectors and eigenvalues of ${\bf L}$, namely, 
\bea
{\bf L}\ \Psi_i = \l_i\ \Psi_i, \hskip 1cm i=1,2,\cdots, {\cal N}. \nonumber
\eea
Let $\psi_{i\a}, \a=1,2,\cdots,{\cal N}$ be the components of $\Psi_i$.
Since  ${\bf L}$ is Hermitian,
 the $\Psi_i$'s  can be taken to be 
 orthonormal satisfying
\bea
( \Psi_i^*, \Psi_j) = \sum_\a \psi_{i\a}^* \psi_{j\a} = \delta_{ij} . \nonumber
\eea
Now the sum of all columns (or rows) of ${\bf L}$ is identically zero, so one of the 
eigenvalues of ${\bf L}$ is identically zero.
It is readily verified that the 
 zero eigenvalue  $\l_1 =0$ 
has the eigenvector
\bea
\x_{1\a} = 1/\sqrt {\cal N}, \hskip 1cm \a=1,2,\cdots,{\cal N}. \nonumber
\eea

We now state our main result as a theorem.

\medskip
\noindent
{\it Theorem: 

\medskip
Consider a resistance network whose Laplacian has nonzero eigenvalues
$\l_i$ with corresponding eigenvectors $\Psi_i = (\psi_{11},\psi_{12},\cdots, \psi_{1{\cal N}})$, 
$i=2,3,\cdots,{\cal N}$. Then
the resistance between nodes $\a$ and $\b$ is given by
 \bea
R_{\a\b} =
\sum_{i=2}^{\cal N} {1\over {\l_i}} \Big| \x_{i\a} - \x_{i\b}\Big|^2.  \label{R}
\eea
}

\medskip\noindent
Proof.

\medskip
We proceed by solving
Eq. (\ref{eq3}) by 
introducing the inverse of the Laplacian ${\bf L}$, or the
  Green's function \cite{lg}.
However, since one of the
eigenvalues of ${\bf L}$ is zero,  the inverse of  ${\bf L}$ must be considered with care.

\medskip
We   add a small term
$\e \, {\bf I}$ to the Laplacian,
where  ${\bf I}$ is the ${\cal N}\times {\cal N}$ identity matrix, 
 and set $\e=0$ at the end.
The modified Laplacian
\bea
{\bf L}(\e) = {\bf L} + \e\, {\bf I} \nonumber
\eea
 is of the same form of    ${\bf L}$ 
except that the diagonal elements $c_i$ are replaced by
$c_i+\e$.  It is clear that ${\bf L}(\e)$ has eigenvalues $\l_i +\e$
and is diagonalized by the same unitary transformation which diagonalizes ${\bf L}$.

\medskip
 The inverse of ${\bf L}(\e)$ is now well-defined and we write
 \bea
{\bf G}(\e) = {\bf L}^{-1}(\e).\nonumber
 \eea
Rewrite (\ref{eq3}) as\, ${\bf L}(\epsilon) {\vec V}(\e) = {\vec I}$
and multiply from the left  by ${\bf G}(\e)$ to obtain         
${\vec V}(\e)={\bf G}(\e)  {\vec I} $.  Explicitly, this reads
 \be
V_i(\e) = \sum_{j=1}^{\cal N} G_{ij}(\e) I_j, \hskip 1cm i=1,2,\cdots,{\cal N}, \label{V}
\ee
where $G_{ij}(\e)$ is the $ij$-th elements of the matrix ${\bf G}(\e)$.
 
\medskip
We now compute the Green's function $G_{ij}(\e)$. 

\medskip
Let ${\bf U}$ be the
unitary matrix which diagonalizes
${\bf L}(\e)$ and ${\bf L}$, namely,
\bea
{\bf U}^\dagger{\bf L}{\bf U} &=& {\bf \Lambda} \nonumber \\
{\bf U}^\dagger{\bf L}(\e){\bf U}&=& {\bf \Lambda}(\e).\label{diag}
\eea
It is readily verified that elements of ${\bf U}$ are
$U_{ij} = \psi_{ji}$, and
 ${\bf \Lambda}$ and  ${\bf \Lambda}(\e)$ are, respectively, diagonal matrices with  elements  
$\l_i\ \d_{ij}$ and $(\l_i+\e)\ \d_{ij}$.

\medskip
 The inverse of the second line of (\ref{diag}) is
\bea
{\bf U}^\dagger {\bf G}(\e){\bf U} = {\bf \Lambda}^{-1}(\e) \nonumber
\eea
where ${\bf \Lambda}^{-1}(\e)$ has elements $(\l_i+\e)^{-1}\d_{ij}$.
It follows that we have
\bea
{\bf G}(\e) = {\bf U} {\bf \Lambda}^{-1}(\e){\bf U}^\dagger \nonumber
\eea
or, explicitly,
 \bea
G_{\a\b} (\e)&=&  \sum_{i=1}^{\cal N} U_{\a\, i}\Bigg(
 \frac 1 {\l_i+\epsilon} \Bigg) U^*_{\b\, i} \nonumber\\
  &=& \frac 1 {{\cal N}\e} + g_{\a\b} (\e) \label{G}
\eea
where
\bea
g_{\a\b} (\e) = \sum_{i=2}^{\cal N}  \frac {\x_{i\a} \x^*_{i\b}} {\l_i+\e}.\label{ggg}
\eea
The substitution of (\ref{G}) into (\ref{V}) now yields, after making use of 
the constraint (\ref{sumrule}),
\bea
V_i(\e) = \sum_{j=1}^{\cal N} g_{ij}(\e) I_j. \nonumber
\eea
It is now safe to take the $\e\to 0$ limit to obtain
\be
V_i = \sum_{j=1}^{\cal N} g_{ij}(0) I_j. \label{VV}
\ee
Finally, by combining (\ref{equalr}), (\ref{I}) with (\ref{VV}) 
 we obtain 
  \bea
R_{\a\b} =
g_{\a\a}(0) +g_{\b\b}(0)
   -g_{\a\b}(0)-g_{\b\a}(0) \nonumber 
 \eea
which becomes (\ref{R}) after introducing (\ref{ggg}). QED \hfill{$\bullet$}
 
\medskip
The usefulness of
(\ref{R}) is illustrated by the following examples.

\medskip
{\bf Example 1}: Consider the 4-node network shown in Fig. 1 with the Laplacian
\bea
{\bf L}=\,\left(\begin{array}{cccc}
 2c_1&-c_1&0&-c_1\\
-c_1&2c_1+c_2&-c_1&-c_2\\
 0&-c_1&2c_1&-c_1 \\
-c_1&-c_2&-c_1&2c_1+c_2\\
\end{array} \nonumber
\right),
\eea
where $c_1=1/r_1, c_2=1/r_2$.  The nonzero eigenvalues of ${\bf L}$ and their
orthonormal eigenvectors are
\bea
\l_2 &=& 4c_1, \hskip1.55cm \Psi_2=\frac 1 2 (1,-1,1,-1) \nonumber \\
\l_3 &=& 2c_1, \hskip1.55cm \Psi_3=\frac 1 {\sqrt 2} (-1,0,1,0) \nonumber \\
\l_4 &=& 2(c_1+c_2), \hskip.5cm \Psi_4=\frac 1 {\sqrt 2} (0,-1,0,1). \nonumber 
\eea
Using (\ref{R}), we obtain 
\bea
R_{13}&=& \frac 1 {\l_2} \Big(\psi_{21}-\psi_{23}\Big)^2+ \frac 1 {\l_3} \Big(\psi_{31}-\psi_{33}\Big)^2
    +\frac 1 {\l_4} \Big(\psi_{41}-\psi_{43}\Big)^2 
  = r_1 ,\nonumber \\
 R_{12}&=& \frac 1 {\l_2} \Big(\psi_{21}-\psi_{22}\Big)^2+ \frac 1 {\l_3} \Big(\psi_{31}-\psi_{32}\Big)^2
    +\frac 1 {\l_4} \Big(\psi_{41}-\psi_{42}\Big)^2
  =\frac {r_1(3r_1+2r_2)} {4(r_1+r_2)} .\nonumber 
\eea

\medskip
{\bf Example 2}: We consider ${\cal N}({\cal N}-1)/2$ 
resistors of equal resistance $r$ on a complete graph of ${\cal N}$ nodes.
A complete graph is a network in which every node is connected to every other node.
The Laplacian is therefore
\be
{\bf L}^{\,\rm complete\>graph }=r^{-1}\,\left(\begin{array}{ccccc}
 {\cal N}-1&-1&\cdots&-1&-1\\
-1&{\cal N}-1&\cdots&-1&-1\\
 \vdots&\vdots&\ddots&\vdots&\vdots\\
-1&-1&\cdots&{\cal N}-1&-1\\
-1&-1&\cdots&-1&{\cal N}-1
\end{array} \label{completegraph}
\right).
\ee
It is readily verified that the Laplacian (\ref{completegraph}) has eigenvalues 
$\lambda_0=0$ and $\lambda_n={\cal N}r^{-1}, n=1,2,...,{\cal N}-1$, 
with corresponding eigenvectors
$\Psi_n$ having components 
\bea
\psi_{n\a} = \frac 1 {\sqrt {\cal N}}\,
  {\rm exp} \,\Bigg( i\,\frac {2\pi n \a}{{\cal N}}\Bigg), \hskip 1cm n,\a = 0,1, ...,
{\cal N}-1.  \nonumber
\eea
It follows from (\ref{R}) the resistance between any two nodes $\a$ and $\b$
is
\bea
R_{\a,\b} &=&  r \, \sum _{n=1}^{{\cal N}} \frac {\Big| \psi _{n\a} -\psi_{n\b} \Big|^2}
  {{\cal N}} = \frac 2 {\cal N}\, r.  \label{compR}
\eea
In subsequent sections we consider applications of (\ref{R}) to regular lattices.

\section{One-dimensional lattice}
It is instructive to first consider the one-dimensional case of  
  a linear array of resistors. We consider  free and
periodic boundary conditions separately.

\medskip\noindent
{\bf Free boundary condition}: 

\medskip
Consider  $N-1$ resistors of resistance $r$ each 
   connected in series forming a chain of $N$ nodes 
 numbered  from $0,1,2,\cdots$, to $N-1$ as shown in Fig. 2,
where each of the two end nodes connects to only one interior node.
This is the Neumann (or the free) boundary condition. 
 The Laplacian (\ref{eq4}) assumes the form 
\bea
{\bf L}^{\,\rm free}_{\{N\times 1\}} = r^{-1} \, {\bf T}_N^{\,\rm free} \nonumber
\eea
where ${\bf T}_N^{\,\rm free}$ is the $N\times N$ matrix
\be
{\bf T}_N^{\,\rm free}=\left(\begin{array}{ccccccc}
1&-1&0&\cdots&0&0&0\\
-1&2&-1&\cdots&0&0&0\\
 \vdots&\vdots&\vdots&\ddots&\vdots&\vdots&\vdots\\
0&0&0&\cdots&-1&2&-1\\
0&0&0&\cdots&0&-1&1
\end{array} \label{TNfree}
\right).
\ee
The  eigenvalues and eigenvectors of ${\bf T}_N$ can be readily computed
(see, for example,  \cite{tw}), and are 
\bea
\lambda_n&=&2\,\Big(1-\cos \tn \Big), \hskip 2.5cm n=0,1,\cdots, N-1\nonumber \\
 \x^{(N)}_{n x}
 &=& \frac 1 {\sqrt N}, \hskip 4.2cm n=0, {\rm \>\>all\>\> }x\nonumber\\
       &=& \sqrt \frac 2 N \cos  \Big((x+  1/ 2) \tn\Big), \hskip 1cm  n=1,2,\cdots,N-1,
 {\rm \>\>all\>\> }x,
\label{psinx}
\eea
where $\tn = n\pi/N$.
 Thus, using (\ref{R}), the resistance between nodes $x_1$ and $x_2$ is
\bea
R^{\,\rm free}_{\{N\times 1\}}(x_1,x_2) &=& \frac r N \sum_{n=1}^{N-1}
\frac {\Big[ \cos \big( x_1+\frac 1 2 \big)\tn - 
\cos \big( x_2+\frac 1 2 \big)\tn \Big]^2 } {1-\cos \tn}\nonumber \\
 &=& r\Big[ F_N(x_1+x_2+1) +F_N(x_1-x_2)\nonumber \\
&& \hskip 1cm -\frac 1 2 F_N(2x_1+1)-\frac 1 2 F_N(2x_2+1) \Big]   ,\label{r13}
\eea
where 
  \bea
 F_N(\ell ) = \frac 1 N \sum_{n=1}^{N-1} \frac {1-\cos (\ell \tn)} {1-\cos \tn}.
 \label{F}
\eea
Note that without loss of generality we can take $0\geq \ell < 2N$.
 
\medskip
The function $F_N(\ell)$ can be evaluated 
by taking the limit of $\l\to 0$ of the function $I_1(0) - I_1(\ell)$ 
evaluated in   (\ref{I1}) below.  It is however instructive to evaluate   
$F_N(\ell)$ directly.  To do this we consider
 the  real part of the summation
\bea
 T_N(\ell)\equiv\frac 1 N \sum_{n=1}^{N-1} \frac {1-e^{i \ell \tn}} {1-e^{i \tn}}. \nonumber
\eea
First, writing out  the real part of the summand, we obtain 
\bea
 {\cal {R}}e\, T_N(\ell) &=&
\frac 1 N \sum_{n=1}^{N-1}
  \frac { 1- \cos \tn - \cos \ell\tn +\cos (\ell-1)\tn } {2(1-\cos\tn)}\nonumber\\ 
&=&\frac 1 2 \Big[ F_N(1)+ F_N(\ell)-F_N(\ell-1)  \Big] . \label{sum3}
\eea
Second, we expand the summand to obtain
\bea
T_N(\ell) =\frac 1 N \sum_{n=1}^{N-1} \sum_{\ell '=0}^{\ell-1} e^{i\pi n\ell'/N}
 \nonumber
\eea
and carry out the summation over $n$.
The term $\ell'=0$ yields $F_N(1)$ and terms  $\ell'\geq 1$ can be explicitly summed,
leading to
 \bea
T_N(\ell) &=& F_N(1) +\frac 1 N \sum_{\ell '=1}^{\ell-1}
 \Bigg[ \frac{1-(-1)^{\ell'}} {1-e^{i\pi \ell'/N}} -1\Bigg],\quad \ell<2N.
  \label{T}
\eea
We now evaluate the real part of $T_N(\ell)$ giving by (\ref{T}). Using the identity
\bea
{\cal R}e\,\Bigg( \frac 1 {1-e^{i\theta}} \Bigg)= \frac 1 2, \hskip1cm
 0<\theta <2\pi ,\label{12}
\eea
we find
  \be
{\cal R}e\, T_N(\ell) = F_N(1)  -\frac 1 {4N} \Big[2\ell -3 -(-1)^\ell\Big]. \label{sum4}
\ee
Equating (\ref{sum3}) with (\ref{sum4}) and noting  $F_N(1) = 1-1/N$, we are led to
the recursion relation
\bea
F_N(\ell)-F_N(\ell-1) = 1 -\frac 1 {2N} \Big[ 2 \ell -1 - (-1)^\ell \Big], \nonumber
\eea
which can be solved (Cf. section 9 below) to yield 
\be
F_N(\ell) = |\ell| - \frac 1 {N} \Bigg( \frac {\ell^2 +|\ell|} 2 
- \Bigg[\frac {|\ell|} 2 \Bigg] \Bigg)
\label{F1}
\ee
where $[x]$ denotes the integral part of $x$.
The substitution of (\ref{F1}) into (\ref{r13}) now gives the answer
\be
R^{\,\rm free}_{\{N\times 1\}}(x_1,x_2)  
  =r\,\big| x_1-x_2\big| \label{1dR}
 \ee  
which is the expected expression.

\medskip\noindent
{\bf Periodic boundary conditions}:

\medskip
Consider next periodic boundary conditions for which nodes $0$ and $N-1$ are also
 connected  as shown in Fig. 3. 
 The  Laplacian ({\ref{eq4}) of the lattice
is therefore ${\bf L}_{\{N\times 1\}}^{\rm per}=r^{-1}\,{\bf T}_{N}^{\rm per}$ where
\bea
{\bf T}_{N}^{\rm per}=\left(\begin{array}{ccccccc}
2&-1&0&\cdots&0&0&-1\\
-1&2&-1&\cdots&0&0&0\\
\vdots&\vdots&\vdots&\ddots&\vdots&\vdots&\vdots\\
0&0&0&\cdots&-1&2&-1\\
-1&0&0&\cdots&0&-1&2
\end{array}
\right) .\label{TNper}
\eea
 The eigenvalues and eigenfunctions of  ${\bf T}_N^{\rm per}$ are well-known and are, respectively,
 \bea
\lambda_n &=& 2\,\Big(1- \cos 2 \tn \Big), \nonumber \\
\psi_{nx}^{\rm per} &=& \frac 1 {\sqrt N}\, 
e^{i2x \tn },  
\hskip 1cm n,x =0,1,\cdots,N-1 .\label{Rper}
\eea
Substituting (\ref{Rper}) into (\ref{R}), we obtain
 the resistance between  nodes $x_1$ and $x_2$ 
\bea
R^{\rm per}_{\{N\times 1\}}(x_1,x_2) &=& \frac r {N} \sum_{n=1}^{N-1}
\frac {\Big| e^{ i 2x_1\tn} - e^{i2x_2\tn}
\Big|^2 } {2\big(1-\cos 2 \tn\big)}\nonumber \\
&=& r\, G_N(x_1-x_2) \label{rr1}
\eea
where 
\bea
G_N(\ell)=
 \frac 1 N \sum_{n=1}^{N-1} \frac {1-\cos (2\ell \tn)} {1-\cos 2\tn}.\nonumber
\eea
 
\medskip
The function $G_N(\ell)$  is again evaluated in a special case of the identity
(\ref{I2}) below.  Following analyses in section 9, one obtains
the recursion relation
 \bea
G_N(\ell)-G_N(\ell-1) = 1- \frac 1 N \big({2\ell -1}\big) \nonumber
\eea
which can be solved  to yield 
\be
G_N(\ell) =
\big|\ell\big| - \ell^2/N. \label{GN}
\ee
 It follows that we have
\be
R^{\rm per}_{\{N\times 1\}}(x_1,x_2) =
r\, \big| x_1-x_2\big| \Bigg[1- \frac {\big| x_1-x_2\big|} N
\Bigg]. \label{R1Dper}
\ee
 The expression (\ref{R1Dper}) is
 the expected resistance of two resistors $\big| x_1-x_2\big|\,r$
and $\big(N-\big| x_1-x_2\big|\big)\,r$  connected in parallel as in a ring.

\section{Two-dimensional network: Free boundaries}
Consider a  rectangular network of resistors  connected in an array of $M\times N$ nodes
forming a network with free boundaries
as shown in Fig. 4.
 
\medskip
Number the nodes by coordinates
 $\{ m,n\}$, 
$0 \leq m \leq M-1 $, $0 \leq  n \leq N-1$ and denote the resistances 
along the two principal directions by
  $r$ and $s$. 
 The Laplacian is  therefore
\be
{\bf L}^{\,\rm free}_{\{M\times N\}} 
= r^{-1}{\bf T}_{M}^{\,\rm free} \otimes {\bf I}_{N}+ s^{-1}
{\bf I}_{M} \otimes {\bf T}_{N}^{\,\rm free}
\ee
where $\otimes$ denotes direct matrix products and ${\bf T}_{N}^{\,\rm free}$ is given by (\ref{TNfree}).
The Laplacian can be diagonalized in  the two subspaces separately, 
yielding eigenvalues
 \be
\lambda_{(m,n)} = 2r^{-1} (1-\cos \tm)  +2s^{-1}(1-\cos \tn), \label{freeeigen}
\ee
and eigenvectors
\be
\x_{(m,n);(x,y)}^{\,\rm free} = \x^{(M)}_{mx}\x^{(N)}_{ny} .
\ee
It then follows from (\ref{R})  that the resistance ${R}_{\,\rm free}$ between two nodes ${\bf r}_1=(x_1, y_1)$
and ${\bf r}_2=(x_2, y_2)$  is 
\bea
&&R_{\{M\times N\}}^{\,\rm free}({\bf r}_1,{\bf r}_2) ={\sum_{m=0}^{M-1}\sum_{n=0}^{N-1}}_{(m,n) \not= (0,0)}
\frac {\Big|\psi_{(m,n);(x_1,y_1)}^{\,\rm free}
 - \psi_{(m,n);(x_2, y_2)}^{\,\rm free}\Big|^2 }{\lambda_{(m,n)}} \nonumber \\
&& = \frac {r} {N} \Big| x_1 -x_2 \Big|  + \frac s {M} \Big| y_1 - y_2 \Big| +\frac 2 {MN} \nonumber\\
&&\times{\sum_{m=1}^{M-1}\sum_{n=1}^{N-1}
\frac {\Big[\cos\Big(x_1+\frac 1 2\Big)\tm \cos\Big(y_1+\frac 1 2\Big)\tn
 - \cos\Big(x_2+\frac 1 2\Big)\tm \cos\Big(y_2+\frac 1 2\Big)\tn
\Big]^2 } 
{r^{-1} (1-\cos \tm )  +s^{-1} (1-\cos \tn )  } } ,\nonumber \\
\label{2dR}
\eea
where
\bea
\tm= \frac{m\pi} M, \hskip1cm \tn= \frac{n\pi} N.\nonumber
\eea 
   Here, use has been made of  (\ref{1dR}) for summing over 
 the $m=0$ and $n=0$ terms.
 The resulting expression (\ref{2dR}) now 
depends on the coordinates $x_1, y_1, x_2$ and $y_2$ individually.

\medskip
The usefulness of (\ref{2dR}) is best illustrated by  applications.
Several examples are now given.

\medskip
{\bf Example 3}:
For  $M=5, N=4, r=s$, 
the resistance between nodes $\{0,0,\}$ and $\{3,3\}$ is computed from (\ref{2dR}) as
\bea
R_{\{5\times 4\}}^{\,\rm free}(\{0,0,\},\{3,3\}) &=& \Bigg( 
\frac 3 4 + \frac 3 5 + \frac {9877231}{27600540}\Bigg)\, r\, \nonumber \\
 &=& (1.70786...)\, r.
\eea

{\bf Example 4}:
For $M=N=4$, we find
\be
R_{\{4\times 4\}}^{\,\rm free}(\{0,0\},\{3,3\}) =\frac {(r+s)(r^2+5rs+s^2)(3r^2+7rs+3s^2)}
{2(2r^2+4rs+s^2)(r^2+4rs+2s^2)}\, .
\ee

{\bf Example 5}: We evaluate the resistance between two nodes in the interior of a large lattice. 
 Consider, for definiteness,
both $M,N=$ odd (for other parities the result (\ref{inffree}) below is the same)
and compute the resistance between two nodes in the center region,
\bea
{\bf r}_1 &=& (x_1,y_1) = \Big( \frac{M-1} 2 +p_1,\, \frac{N-1} 2 +q_1\Big) \nonumber\\
{\bf r}_2 &=& (x_2,y_2) = \Big( \frac{M-1} 2 +p_2,\, \frac{N-1} 2 +q_2\Big), \nonumber
\eea
where  $p_i,q_i<<M,N$ are integers.  The numerator of the summand in (\ref{2dR}) becomes
\bea
 \Big[\cos \Big(\frac {m\pi} 2&+& p_1 \tm\Big) \cos \Big(\frac {n\pi} 2 + q_1 \tn\Big)
-\cos \Big(\frac {m\pi} 2 + p_2 \tm\Big) \cos \Big(\frac {n\pi} 2 + q_2 \tn\Big)\Big]^2
 \nonumber \\
&=& \Big(\cos  p_1 \tm \cos  q_1 \tn - \cos  p_2 \tm \cos  q_2 \tn\Big)^2,  
 \quad m,n = {\rm even} \nonumber \\
&=& \Big(\sin  p_1 \tm \,\sin q_1 \tn - \sin p_2 \tm \, \sin q_2 \tn\Big)^2,  
 \quad m,n = {\rm odd} 
\nonumber \\
&=& \Big(\sin  p_1 \tm \,\cos q_1 \tn - \sin p_2 \tm \, \cos q_2 \tn\Big)^2,  
 \quad m= {\rm odd}, n={\rm even} \nonumber \\
&=& \Big(\cos  p _1 \tm \,\sin q_1 \tn - \cos p_2 \tm \, \sin q_2 \tn\Big)^2,  
 \quad m= {\rm even} , n={\rm odd}\nonumber
 \eea 
and the summation in (\ref{2dR}) breaks 
into four parts.  In the $M,N\to\infty$ limit
the summations can be replaced by integrals.
After some reduction we arrive at the expression
\be
R_{\,\rm \infty}({\bf r}_1,{\bf r}_2)
= \frac 1 {(2\pi)^2} \int_0^{2\pi}d\t \int_0^{2\pi}d\p \Bigg(
\frac {1- \cos \Big(x_1-x_2)\p\, \cos(y_1-y_2)\t  } 
 {r^{-1}\big(1-\cos \p\big)+s^{-1}\big(1-\cos \t \big)}\Bigg). \label{inffree}
\ee
 Our expression (\ref{inffree}) agrees with 
 the known expression obtained previously  \cite{cserti}.  It can  be verified
 that the expression (\ref{inffree}) holds between any two nodes in the lattice, 
provided that  the two nodes are far from the boundaries.
 
\section{Two-dimensional network: periodic boundary conditions}
We next consider  an $M\times N$ network with   periodic boundary conditions.
 The Laplacian in this case is
\be
{\bf L}^{\rm per}_{\{M\times N\}}
 =  r^{-1}{\bf T}_{M}^{\rm per} \otimes {\bf I}_{N}+  s^{-1} {\bf I}_{M} \otimes {\bf T}_{N}^{\rm per}
\label{2dperL}
\ee
 where ${\bf T}_{N}^{\rm per}$ is given by (\ref{TNper}).
  The Laplacian (\ref{2dperL}) can again be 
diagonalized in the two subspace separately, yielding
eigenvalues and eigenvectors 
 \bea
\lambda_{(m,n)} &=& 2r^{-1} (1-\cos 2\tm)  +2s^{-s} (1-\cos 2 \tn)\nonumber \\ 
\psi_{(m,n);(x,y)} &=& \frac 1 {\sqrt {MN}}\, e^{i2x\tm} e^{i2y\tn}.
   \label{pereigen}
\eea
This leads to the resistance 
 between  nodes ${\bf r}_1=(x_1, y_1)$
and  ${\bf r}_2=(x_2, y_2)$  
 \bea
R_{\{M\times N\}}^{\rm per} ({\bf r}_1,{\bf r}_2) 
&=&  {\sum_{m=0}^{M-1}\sum_{n=0}^{N-1}} _{(m,n) \not= (0,0)}
\frac {\Big|\psi_{(m,n);(x_1,y_1)} - \psi_{(m,n);(x_2, y_2)}\Big|^2 }{\lambda_{(m,n)}} \nonumber \\
&=& \frac  r N \Bigg[\,\Big| x_1 -x_2 \Big| -\frac {(x_1-x_2)^2} M
\Bigg] + \frac  sM \Bigg[\,\Big| y_1 -y_2 \Big| -\frac {(y_1-y_2)^2} N
\Bigg] \nonumber \\
&+& \frac 1 {MN} \sum_{m=1}^M \sum_{n=1}^N
\frac {1- \cos \Big[2(x_1-x_2)\tm+2(y_1-y_2)\tn \Big] } 
 {r^{-1}\big(1-\cos 2\tm \big)+s^{-1}\big(1-\cos 2\tn \big)} ,\nonumber \\
&& \hskip 2cm \label{RR}          
\eea
where  the two terms in the second line are given by (\ref{R1Dper}).
It is clear that the result depends only on the differences
$\big|x_1-x_2\big|$ and $\big|y_1-y_2\big|$, as it should 
under periodic boundary conditions.

\medskip
{\bf Example 6}: Using (\ref{RR})  the resistance
between nodes $\{0,0\}$ and $\{3,3\}$ on  a $5\times 4$ periodic lattice
with $r=s$ is
\bea
R^{\rm per}_{\{5\times 4\}}\Big(\{0,0\},\{3,3\}\Big) &=& \bigg(\frac 3 {10}
+ \frac 3 {20} +\frac{1799}{7790}\Bigg) r \nonumber \\
&=& (0.680937...)\, r. \label{rper}
\eea
This is to be compared to the value \ $1.707863...\  r$ \ for free boundary conditions
given in Example 3.   It can also be verified that the  resistance
between nodes $(\{0,0\})$ and $(\{2,1\})$ is also given by (\ref{rper})
as it must for a periodic lattice.

\medskip
In the limit of $M,N\to\infty$ with $\big|{\bf r}_1-{\bf r}_2\big|$ finite, 
(\ref{RR}) becomes
\bea
R_{\rm \infty} ({\bf r}_1,{\bf r}_2)
&=&  \frac 1 {(2\pi)^2} \int_0^{2\pi}d\t \int_0^{2\pi}d\p 
\frac {1- \cos \Big[(x_1-x_2)\p+(y_1-y_2)\t \Big] } 
 {r^{-1}\big(1-\cos \p\big)+s^{-1}\big(1-\cos \t \big)} \nonumber \\
&=&  \frac 1 {(2\pi)^2} \int_0^{2\pi}d\t \int_0^{2\pi}d\p 
\frac {1- \cos (x_1-x_2)\p\cos(y_1-y_2)\t  } 
 {r^{-1}_1\big(1-\cos \p\big)+s^{-1}\big(1-\cos \t \big)} ,\label{infper}        
\eea
which agrees with  (\ref{inffree}).

\section{Cylindrical boundary conditions}
Consider an $M\times N$ resistor network embedded on a cylinder
with periodic boundary in the  direction of $M$ and  free boundaries
 in the direction of $N$.

\medskip
The Laplacian is  
\bea
{\bf L}^{\rm cyl}_{\{M\times N\}} 
=  r^{-1}\,{\bf T}_{M}^{\rm per} \otimes {\bf I}_{N}+  s^{-1}\,{\bf I}_{M} \otimes {\bf T}_{N}^{\,\rm free}
\nonumber
\eea
which can again be diagonalized in the two subspaces separately. 
This gives the eigenvalues and eigenvectors
 \bea
\lambda_{(m,n)} &=& 2r^{-1} (1-\cos 2\tm)  +2s^{-1} (1-\cos \tn), \nonumber \\
\x_{(m,n);(x,y)}^{\rm cyl} &=& \frac 1 {\sqrt M}\, e^{i 2x\tm}\x^{(N)}_{ny} . \nonumber
\eea
It follows that the resistance ${R}_{\,\rm free}$ between nodes ${\bf r}_1=(x_1, y_1)$
and ${\bf r}_2=(x_2, y_2)$  is 
\bea
&&R_{\{M\times N\}}^{\rm cyl}({\bf r}_1,{\bf r}_2) ={\sum_{m=0}^{M-1}\sum_{n=0}^{N-1}}_{(m,n) \not= (0,0)}
\frac {\Big|\psi_{(m,n);(x_1,y_1)}^{\rm cyl}
 - \psi_{(m,n);(x_2, y_2)}^{\rm cyl}\Big|^2 }{\lambda_{(m,n)}} \nonumber \\
&& =\frac  r N \Bigg[\,\Big| x_1 -x_2 \Big| -\frac {(x_1-x_2)^2} M
\Bigg] + \frac s {M} \Big| y_1 - y_2 \Big|  \nonumber\\
&&+\frac 2 {MN}\,{\sum_{m=1}^{M-1}\sum_{n=1}^{N-1}
\Bigg(\frac {C_1\,^2+C_2\,^2 
 - 2C_1C_2 \cos 2(x_1-x_2)\tm } 
{r^{-1} (1-\cos 2\tm )  +s^{-1} (1-\cos \tn )  } }\Bigg),\nonumber 
\label{cylR}
\eea
where 
\bea
C_1 = \cos\Big(y_1+\frac 1 2\Big)\tn,
\hskip1cm  C_2=\cos\Big(y_2+\frac 1 2\Big)\tn. \label{CC}
\eea
 It can be verified that in the $M,N \to \infty$ limit
(\ref{cylR}) leads to the same expression (\ref{inffree}) for 
 two interior nodes in an infinite lattice.

\medskip
{\bf Example 7}: The resistance
between nodes $\{0,0\}$ and $\{3,3\}$ on  a $5\times 4$ cylindrical lattice
with $r=s$ is computed to be
\bea
R^{\rm cyl}_{\{5\times 4\}}\Big(\{0,0\},\{3,3\}\Big) &=& \bigg(\frac 3 {10}
+ \frac 3 5 +
\frac{5023}{8835}\Bigg) r \nonumber \\
&=& (1.46853...)\, r.
\eea
This is compared to the values of \ $(1.70786...)\,  r$ \ for free boundary conditions
and \ $ (0.680937...)\, r$\  for periodic boundary conditions.

\section{M\"obius strip}
We next consider an $M\times N$ resistor lattice embedded on a
M\"obius strip of width $N$  and length $M$, which is a rectangular strip connected
at two ends after a $180^o$ twist of one of the two ends of the strip.  
The schematic figure of a M\"obius strip is shown in Fig. 5(a).
 The Laplacian for
this lattice assumes the form
\bea
{\bf L}^{\rm Mob}_{\{M\times N\}} 
=  r^{-1}\Big[{\bf H_M} \otimes {\bf I}_{N}
-{\bf K}_M \otimes {\bf J}_N \Big]+  s^{-1} {\bf I}_{M} \otimes {\bf T}_{N}^{\,\rm free}
\label{mobLaplacian}
\eea
where
\bea
{\bf K}_N&=&\left(\begin{array}{ccccc}
0&0&\cdots&0&1\\
0&0&\cdots&0&0\\
 \vdots&\vdots&\ddots&\vdots&\vdots\\
0&0&\cdots&0&0\\
1&0&\cdots&0&0
\end{array}\right), 
\quad
{\bf J}_N=\left(\begin{array}{ccccc}
0&0&\cdots&0&1\\
0&0&\cdots&1&0\\
 \vdots&\vdots&\ddots&\vdots&\vdots\\
0&1&\cdots&0&0\\
1&0&\cdots&0&0
\end{array}\right), \nonumber \\
 {\bf H}_N &=& {\bf T}^{\rm per}_N+{\bf K}_N =  \left(\begin{array}{ccccc}
2&-1&\cdots&0&0\\
-1&2&\cdots&0&0\\
 \vdots&\vdots&\ddots&\vdots&\vdots\\
0&0&\cdots&2&-1\\
0&0&\cdots&-1&2
\end{array}\right) \nonumber 
\eea
are $N\times N$ matrices. 
Now $ {\bf T}_{N}^{\,\rm free}$ and ${\bf J}_N$ commute so they can be replaced by their
respective eigenvalues $2(1-\cos \tn)$ and $(-1)^n$ and we need only to diagonalize
an $M\times M$ matrix.  This leads to the following
 eigenvalues
and eigenvectors of the Laplacian (\ref{mobLaplacian})  \cite{tw,tzeng}:
\bea
\lambda_{(m,n)}&=& 2r^{-1}\cos \Big[\Big( {4m+1-(-1)^n}\Big)  \frac \pi {2M}\Big]+2s^{-1}
\Big(1-\cos\frac {n\pi} N \Big), \nonumber\\
\psi^{\rm Mob} _{(m,n);(x,y)} &=& \frac 1 {\sqrt M}\, 
{\rm exp} \Big[ {i\Big(4m+1-(-1)^n\Big) \frac {x\pi} {2M}}\Big] \cdot \psi^{(N)}_{ny}\label{mob}
 \eea
where $\x^{(N)}_{n y}$ is given in (\ref{psinx}).
Substituting these expressions into (\ref{R}) and after a little reduction, we obtain
\bea
 &&R_{\{M\times N\}}^{\rm Mob}({\bf r}_1,{\bf r}_2) ={\sum_{m=0}^{M-1}\sum_{n=0}^{N-1}}_{(m,n) \not= (0,0)}
\frac {\Big|\psi^{\rm Mob}_{(m,n);(x_1,y_1)}
 - \psi^{\rm Mob}_{(m,n);(x_2, y_2)}\Big|^2 }{\lambda_{(m,n)}} \nonumber \\
 &&\hskip 2.8cm = \frac  r N \Bigg[\,\Big| x_1 -x_2 \Big| -\frac {(x_1-x_2)^2} M
\Bigg] \nonumber\\
&&+\frac 1 {MN}\,\sum_{m=0}^{M-1}\sum_{n=1}^{N-1}
\frac {C_1\,^2+C_2\,^2 
 - 2C_1C_2 \cos\Bigg[ (x_1-x_2)\big(4m+1-(-1)^n\big) \frac \pi {2M}\Bigg] } 
{r^{-1}\Big[1-\cos\big(4m+1-(-1)^n\big) \frac \pi {2M}\Big] +s^{-1} (1-\cos \tn )  } ,\nonumber \\
\label{mobR}
\eea
where $C_1$ and $C_2$ have been given in (\ref{CC}).

\medskip
{\bf Example 8}: The $2\times 2$ M\"obius strip is a complete graph of ${\cal N}=4$ nodes.
For $r=s$ the expression ({\ref{mobR}) gives a resistance  $r/2$ between any two nodes which
agrees with (\ref{compR}).

\medskip
{\bf Example 9}: The resistance
between nodes $(0,0)$ and $(3,3)$ on  a $5\times 4$ M\"obius strip with $r=s$ is
computed from (\ref{mobR}) as
\bea
R^{\rm Mob}_{\{5\times 4\}}\Big(\{0,0\},\{3,3\}\Big) &=& \bigg( \frac 3 {10} +
\frac{1609}{2698}\Bigg) r \nonumber \\
&=& (0.896367...)\, r.
\eea
This is to be
compared to the corresponding values 
for the same network under other boundary conditions in Examples $3,6,$ and 7.  
 
\section{Klein bottle}
A Klein bottle is a M\"obius strip with a periodic boundary condition imposed in the  
other direction.  We consider an $M\times N$ resistor grid embedded on a Klein bottle,
a schematic figure of which is shown in Fig. 5(b).

\medskip
Let the network have a twisted boundary condition in the direction of the length $M$ 
and a periodic boundary condition in the direction of the width $N$.
Then, in analogous to (\ref{mobLaplacian}), the Laplacian of the network assumes the form
\bea
{\bf L}^{\rm Klein}_{\{M\times N\}} 
=  r^{-1}\Big[{\bf H_M} \otimes {\bf I}_{N}
-{\bf K}_M \otimes {\bf J}_N \Big]+  s^{-1} {\bf I}_{M} \otimes {\bf T}_{N}^{\,\rm per}.
\label{kleinLaplacian}
\eea

Now the matrices ${\bf J}_N$ and ${\bf T}_N^{\,\rm per}$ commute so they can be replaced
by their respective eigenvalues $\pm 1$ and $2(1-\cos 2\tn)$ in (\ref{kleinLaplacian})
and one needs only to diagonalize an $M\times M$ matrix.
This leads to the following eigenvalues and eigenvectors for ${\bf L}^{\rm Klein}_{\{M\times N\}}$
\cite{tw,tzeng}:
\bea
\lambda_{(m,n)}(\tau)&=& 2r^{-1}\Bigg[1-\cos \Big((2m+\tau)  \frac \pi {M}\Big)\Bigg]+2s^{-1}
\Big(1-\cos\frac {2n\pi} N \Big), \nonumber\\
\psi^{\rm Klein} _{(m,n);(x,y)} &=& \frac 1 {\sqrt M}\, 
{\rm exp} \Big[ {i\Big(2m+\tau\Big) \frac {x\pi} {M}}\Big] \cdot \psi^{(N)\dagger}_{ny},
 \label{kleineigen}
\eea
where
\bea
\tau=\tau_n&=& 0, \hskip1cm n=0,1,\cdots, \Bigg[\frac{N-1} 2 \Bigg], \nonumber \\
      &=& 1, \hskip1cm n= \Bigg[\frac{N+1} 2 \Bigg], \cdots, N-1, \nonumber 
\eea
and
\bea
\psi^{(N)\dagger}_{ny}  &=& \frac 1 {\sqrt N}, \hskip 4cm n=0, \nonumber \\
   &=& \sqrt {\frac 2 N} \cos \Big[(2y+1)\frac {n\pi} N\Big], \hskip1cm n=1,2,\cdots, 
 \Bigg[\frac{N-1} 2 \Bigg], \nonumber \\
   &=& \frac 1 {\sqrt N}(-1)^y \hskip 3.3cm n= \frac N 2, \hskip 0.5cm 
            {\rm for\>\>even\>\>}N {\rm \>\> only}, \nonumber \\
   &=& \sqrt {\frac 2 N} \sin \Big[(2y+1)\frac {n\pi} N\Big], \hskip1.2cm n=\Bigg[\frac{N} 2 
     \Bigg]+1, \cdots, N-1. \nonumber 
\eea
 Substituting these expressions in (\ref{R}), separating out the summation for $n=0$,
and making use of the identity
\bea
\sin \Big[(2y+1)\Big( \frac N 2 +n\Big) \frac \pi N \Big] =(-1)^y \cos \Big[(2y+1)
\frac {n\pi} N \Big] , \nonumber 
\eea
we obtain after some reduction
\bea
 &&R_{\{M\times N\}}^{\rm Klein}({\bf r}_1,{\bf r}_2) ={\sum_{m=0}^{M-1}\sum_{n=0}^{N-1}}_{(m,n) \not= (0,0)}
\frac {\Big|\psi^{\rm Klein}_{(m,n);(x_1,y_1)}
 - \psi^{\rm Klein}_{(m,n);(x_2, y_2)}\Big|^2 }{\lambda_{(m,n)}(\tau_n)} \nonumber \\ 
&& \quad = \frac  r N \Bigg[\,\Big| x_1 -x_2 \Big| -\frac {(x_1-x_2)^2} M
\Bigg]  \nonumber\\
&&\quad \quad +\,\,
\sum_{m=0}^{M-1}\sum_{n=1}^{N-1}
\frac 1 {\l_{(m,n)}(\tau_n) } \Big|  \psi^{\rm Klein} _{(m,n);(x_1,y_1)}
 -  \psi^{\rm Klein} _{(m,n);(x_2,y_2)} \Big|^2  \nonumber \\
&& \quad = \frac  r N \Bigg[\,\Big| x_1 -x_2 \Big| -\frac {(x_1-x_2)^2} M
\Bigg] + \Delta_N \nonumber\\
&&\quad \quad +\,\frac 2 {MN}\,\sum_{\tau=0}^1 \sum_{m=0}^{M-1}\sum_{n=1}^{[\frac {N-1} 2]}
\frac {C_1^2+C_2^2-2(-1)^{(y_1-y_2)\tau}C_1C_2 \cos\big[\,2(x_1-x_2) \Theta_m(\tau)\,\big]}
 {\l_{(m,n)}(\tau) }, \nonumber \\
 \label{kleinR}
\eea
where
\bea
\Theta_m(\tau) &=& \Big(m +  \frac {\tau }{2}\Big) \frac \pi M \ ,\nonumber \\
\Delta_N &=& \frac 2 {MN} \sum_{m=0}^{M-1} 
   \frac { 1-(-1) ^{y_1-y_2} \cos\big[ \,2(x_1-x_2) \Theta_m(1)\,\big]}
  {\lambda_{(m,N/2)}(1)},  \quad N = {\rm even} \nonumber \\
  &=& 0, \hskip 1cm N = {\rm odd},
\eea
and $C_i= \cos [(y_i+1/2) n\pi/N], i=1,2,$  as defined in (\ref{CC}). 
  
\medskip
{\bf Example 10}: The resistance
between nodes $(0,0)$ and $(3,3)$ on  a $5\times 4$ ($N=$ even) Klein bottle with $r=s$ is
computed from (\ref{kleinR}) as
\bea
R^{\rm Klein}_{\{5\times 4\}}\Big(\{0,0\},\{3,3\}\Big) &=& \bigg( \frac 3 {10} + \frac 5 {58}
+\frac{56}{209}\Bigg) r \nonumber \\
&=& (0.654149...)\, r,
\eea
where the three terms in the first line are from the evaluation of 
corresponding terms in (\ref{kleinR}).  The result is to be
compared to the corresponding value
for the same $5\times 4$ network under the M\"obius boundary condition considered in Example 9,
which is the Klein bottle without periodic boundary connections.

\section{Higher-dimensional lattices}
The two-point resistance can be computed using (\ref{R}) for lattices
in any spatial dimensionality under various boundary conditions.
  To illustrate,
we give the result for an $M\times N\times L$ cubic 
lattice with free boundary conditions.

\medskip
 Number the nodes  by $\{ m,n.\ell\}$, 
$0 \leq m \leq M-1 $, $0 \leq  n \leq N-1, 0 \leq \ell \leq L-1$, and let  the resistances 
along the  principal axes be, respectively,
  $r$, $s$, and $t$. 
 The Laplacian then assumes the form
\bea
{\bf L}^{\,\rm free}_{\{M\times N\times L\}} 
= r^{-1}{\bf T}_{M}^{\,\rm free} \otimes {\bf I}_{N}\otimes {\bf I}_L+ s^{-1}
{\bf I}_{M} \otimes   {\bf T}_{N}^{\,\rm free} \otimes {\bf I}_L 
+ t^{-1}{\bf I}_{M} \otimes   {\bf I}_{N} \otimes   {\bf T}_{L}^{\,\rm free} \nonumber
 \eea
where ${\bf T}_{N}^{\,\rm free}$ is given by (\ref{TNfree}).
The Laplacian can be diagonalized in  the three subspaces separately, 
yielding eigenvalues
\be\lambda_{(m,n,\ell)} =2\,r^{-1} (1-\cos \tm) +2\,s^{-1} (1-\cos \tn) +2
\,t^{-1} (1-\cos\a_\ell), \label{3deigen}
\ee
and eigenvectors
\bea
\x_{(m,n,\ell);(x,y,z)}^{\,\rm free} = \x^{(M)}_{mx}\x^{(N)}_{ny}\x^{(L)}_{\ell z} \nonumber
\eea
where $\x^{(M)}_{mx}$ is given by (\ref{psinx})  and $\a_\ell = \ell\pi/L$.  
It then follows from (\ref{R})  that the resistance ${R}_{\,\rm free}$ between two nodes 
${\bf r}_1=(x_1, y_1,z_1)$
and ${\bf r}_2=(x_2, y_2,z_2)$  is 
\bea
R_{\{M\times N\times L\}}^{\,\rm free}({\bf r}_1,{\bf r}_2) 
 &=&{\sum_{m=0}^{M-1}\sum_{n=0}^{N-1}}{\sum_{\ell=0}^{L-1}}_{(m,n,\ell) \not= (0,0,0)}
\lambda_{(m,n,\ell)}^{-1} \nonumber \\
 && \times\, {\Big|\psi_{(m,n,\ell);(x_1,y_1,z_1)}^{\,\rm free}
 - \psi_{(m,n,\ell);(x_2, y_2,z_2)}^{\,\rm free}\Big|^2 }
\eea
The summation can be broken down as
\bea
&&R_{\{M\times N\times L\}}^{\,\rm free}({\bf r}_1,{\bf r}_2) 
={\sum_{m=1}^{M-1}\sum_{n=1}^{N-1}}{\sum_{\ell=1}^{L-1}}
\frac {\Big|\psi_{(m,n,\ell);(x_1,y_1,z_1)}^{\,\rm free}
 - \psi_{(m,n,\ell);(x_2, y_2,z_2)}^{\,\rm free}\Big|^2 }{\lambda_{(m,n,\ell)}} \nonumber \\
&& +
\frac 1 L R_{\{M\times N\}}^{\,\rm free}\Big(\{x_1,y_1\},\{x_2,y_2\}\Big) 
 + \frac 1 M R_{\{N\times L\}}^{\,\rm free}\Big(\{y_1,z_1\},\{y_2,z_2\}\Big)\nonumber \\ 
&&
+ \frac 1 N R_{\{ L\times M\}}^{\,\rm free}\Big(\{z_1,x_1\},\{z_2,x_2\}\Big) \nonumber \\
&&- \frac 1 {MN} R_{\{L\times 1\}}^{\,\rm free}\big(x_1,x_2\big)
- \frac 1 {NL} R_{\{M\times 1\}}^{\,\rm free}\big(y_1,y_2\big) 
- \frac 1 {LM} R_{\{N\times 1\}}^{\,\rm free}\big(z_1,z_2\big). \label{3dR}
 \eea
All terms in (\ref{3dR}) have previously been computed except the summation in the first 
line.
 
\medskip
{\bf Example 11}: The resistance between the nodes $(0,0,0)$ and $(3,3,3)$ in a
$5\times 5\times 4$ lattice with free boundaries  and $r=s=t$ is computed from (\ref{3dR})
as
\bea
R_{\{5\times 5\times 4\}}^{\,\rm free}\big( \{0,0,0\};\{3,3,3\} \big) &=&\Bigg(
\frac {327687658482872}{352468567489225} \Bigg) \,r \nonumber \\
&=& (0.929693...)\, r.
\eea

\medskip
{\bf Example 12}:  The resistance between two interior nodes ${\bf r}_1$ and $ {\bf r}_2$
 can be worked out as in Example 4.
The result is 
\bea
&&R_{\,\rm \infty}({\bf r}_1,{\bf r}_2)
= \frac 1 {(2\pi)^3} 
 \int_0^{2\pi}d\t \int_0^{2\pi}d\p\int_0^{2\pi}d\a  \nonumber \\
  && \quad \times \Bigg(
\frac {1- \cos \Big(x_1-x_2)\p\, \cos(y_1-y_2)\t\cos(z_1-z_2)\a  } 
 {r^{-1}\big(1-\cos \p\big)+s^{-1}\big(1-\cos \t \big)+t^{-1}\big(1-\cos \a \big)}\Bigg), \nonumber
 \eea
which is the known result \cite{cserti}.

\section{Summation and product identities}
The  reduction of the two-point resistances for one-dimensional lattices
to the simple and familiar
expressions of (\ref{rr1}) and (\ref{R1Dper})
is facilitated by the use of the summation identities (\ref{F1}) and(\ref{GN}).
In this section we extend the consideration and 
 generalize these identities which can be used to reduce the computational labor for
lattice sums as well as analyze large-size expansions in two-and-higher dimensions.
 
\medskip
We state two new lattice sum identities as a Proposition.
 
\medskip
\noindent
{\it Proposition: Define 
\bea
I_\a (\ell) =  \frac 1 N \sum_{n=0}^{N-1} \frac {\cos\big(\a\,\ell\, \frac {n\pi} N \big)} 
  {\cosh \l -\cos \big(\a \,\frac {n\pi} N \big)}, \hskip1cm \a=1,2. \nonumber
\eea
Then the following identities hold for $\l \geq 0$, $N=1,2,\cdots,$ }
 \bea 
 I_1(\ell)  &=&\frac {\cosh(N-\ell)\l}{(\sinh \l) \sinh (N\l)}  + \frac 1 N \Bigg[ \frac 1 {\sinh^2\l} 
  + \frac {1-(-1)^\ell} {4\cosh^2(\l/2)}\Bigg],\ 0 \leq \ell< 2N,\nonumber \\
  \label{I1} \\
 I_2(\ell) &=&\frac {\cosh \big(\frac N 2 - \ell\big) \l} {(\sinh \l) \sinh (N\l/2)}\, , 
          \hskip 4.2cm 0 \leq \ell< N   .   \label{I2} 
 \eea
\noindent
Remarks:

\medskip
1. It is clear that without the loss of generality we can restrict $\ell$ to the ranges indicated.

\medskip
2. For $\ell=0$ and $\l\to 0$, $I_1(0)$ leads to  (\ref{F1}) and  $I_2(0)$  leads to (\ref{GN}). 

\medskip
3. In the  $N\to\infty$ limit both  (\ref{I1}) and  (\ref{I2}) become the  integral
 \be
\frac 1 \pi \int_0^\pi \frac {\cos (\ell \p)} {\cosh \l - \cos \p}  d\,\p
= \frac {e^{-\ell \l}} {\sinh \l}  \quad \quad \ell \geq 0 .
\ee

\medskip
4.  Set $\ell = 0$ in (\ref{I1}), multiply by $\sinh \l$ and integrate over $\l$, we obtain the product 
identity 
\be
\prod_{n=0}^{N-1} \bigg( \cosh \l -\cos \frac {n\pi} N \bigg) = (\sinh N \l) \tanh ( {\l}/ 2).
\ee

\medskip
5. Set $\ell = 0$ in (\ref{I2}), multiply by $\sinh \l$ and integrate over $\l$.  We obtain the product 
identity
\be
\prod_{n=0}^{N-1} \bigg( \cosh \l -\cos \frac {2n\pi} N \bigg) = \sinh^2 ( {N\l}/ 2).
\ee

\medskip
\noindent
 Proof of the proposition:

\medskip
It is convenient to introduce the notation
\bea
S_\a(\ell) = 
\frac 1 N \sum_{n=0}^{N-1} \frac {\cos(\ell\, \p_n)} 
  {1+a^2 -2a\cos \p_n}, \hskip 1cm a<1 , \quad \a=1,2 \label{SL}
\eea
so that 
\be
I_\a(\ell) = 2\, a\, S_\a(\ell), \hskip1cm  a=e^{-\l}.
\ee
 It is readily seen that we have the identity
\be 
S_\a(1) = \frac 1{2a} \Big[ (1+a^2) S_\a (0)-1\Big]. \label{S10}
\ee

\medskip
\noindent
1. Proof  of (\ref{I1}):

\medskip
First we evaluate $S_1(0)$ by carrying
 out the following summation, where ${\cal R}$e denotes the real part, 
in two different ways.  First we have
\bea
{\cal {R}}e\, 
\frac 1 N \sum_{n=0}^{N-1}
  \frac 1 { 1- a\, e^{i\p_n}}
 &=& {\cal {R}}e\, \frac 1 N \sum_{n=0}^{N-1}
  \frac {1-a \, e^{-i\p_n}} {\Big| 1- a\, e^{i\p_n}\Big|^2} \nonumber \\
&=& \frac 1 N \sum_{n=0}^{N-1} \frac {1-a \cos \p_n}{1+a^2-2a\cos \p_n} \nonumber \\
&=& S_1(0) -a S_1(1) \nonumber \\
&=& \frac 1 2 \Big[1+ (1-a^2)  S_1(0)\Big].  \label{I11}
\eea
Secondly by expanding the summand we have 
\bea
{\cal {R}}e\,\frac 1 N \sum_{n=0}^{N-1}
  \frac 1 { 1- a\, e^{i\p_n}}
={\cal {R}}e\,\frac 1 N \sum_{n=0}^{N-1} \sum_{\ell=0}^\infty a^\ell e^{i\ell n\pi/N}
 \nonumber
\eea
and carry out the summation over $n$ 
for fixed $\ell$.   
  It is clear that all $\ell=$ even terms vanish except those with $\ell = 2 m N, m=0,1,2,...$ 
which yield $\sum_{m=0}^\infty a^{2mN} =1/(1-a^{2N})$.
For $\ell = $ odd $=2m+1,$ $m=0,1,2,...$ we have 
\bea
{\cal {R}}e\, \sum_{n=0}^{N-1} e^{i(2m+1) n\pi/N} = {\cal {R}}e\,\frac {1-(-1)^{2m+1}}
{1-e^{i(2m+1)\pi/N}} = 1 \nonumber 
\eea
after making use of (\ref{12}).  
So the summation over $\ell=$ odd terms  yields
$N^{-1}\sum_{m=0}^\infty a^{2m+1} =a/N(1-a^{2})$, and  we have
\be
{\cal {R}}e\, \sum_{n=0}^{N-1}\frac 1 { 1- a\, e^{i\p_n}}
   =  \frac 1 {1-a^{2N}} +\frac a {N(1-a^2)} \label{I12}
\ee
Equating (\ref{I11}) with  (\ref{I12})  we obtain 
\be
S_1(0) = \frac 1 {1-a^2} \Bigg[ \Bigg(\frac {1+a^{2N}} {1-a^{2N}}\Bigg)
  + \frac {2a} {N(1-a^2)} \Bigg] .\label{SN1}
\ee
 
\medskip
To evaluate $S_1(\ell)$ for general $\ell$,  we
consider the summation
 \bea
{\cal {R}}e\, 
\frac {1} N \sum_{n=0}^{N-1}
  \frac {1-\big(a\, e^{i \p_n}\big)^\ell} { 1- a\, e^{i \p_n}}
 &=& {\cal {R}}e\, 
\frac {1} N \sum_{n=0}^{N-1}
  \frac {(1-a^\ell\, e^{i \ell \p_n})(1-a\, e^{-i\p_n})} {\big|1- a\, e^{i \p_n}\big|^2}\nonumber \\
&=&  S_1(0) - a S_1(1) - a^\ell S_1(\ell) +a^{\ell+1} S_1(\ell-1),\label{I41}
\eea
where the second line is obtained by writing out  the real part of the summand as in (\ref{I11}).
 On the other hand, by expanding the summand we have
\bea
{\cal {R}}e\, 
\frac {1} N \sum_{n=0}^{N-1}
  \frac {1-\big(a\, e^{i \p_n}\big)^\ell} { 1- a\, e^{i \p_n}} &=&
{\cal {R}}e\, 
\frac {1} N \sum_{n=0}^{N-1} \sum_{m=0}^{\ell -1} a^m e^{i\pi m  n/N} \nonumber \\
&=& 1+{\cal {R}}e\, 
\frac {1} N \sum_{m=1}^{\ell-1} a^m \Bigg(\frac {1-(-1)^m} {1-e^{i\pi m/N}} \Bigg) \nonumber \\
&=& 1+\frac {a(1-a^\ell)} {N(1-a^2)}, \hskip 1cm \ell = {\rm even}< 2N \nonumber \\
&=& 1+\frac {a(1-a^{\ell-1})} {N(1-a^2)}, \hskip .9cm \ell = {\rm odd} <2N ,\label{I42}
\eea
where again we have used (\ref{12}).
   
\medskip
 Equating (\ref{I42}) with (\ref{I41}) and using 
({\ref{S10})   and (\ref{SN1}),
 we obtain the recursion relation
\bea
S_N(\ell)-a\,S_N(\ell-1)=A\,a^{-\ell} +B_\ell \label{Q}
\eea
 where 
\bea
A =\frac {a^{2N}} {1- {a^{2N}} }, \hskip 1cm B_\ell = \frac {a^{(1+(-1)^\ell)/2} }{ N(1-a^2)} .
\eea
 
The recursion relation (\ref{Q}) can be solved by standard means.  Define
 the generating function
\be
G_\a(t) =\sum_{\ell =0}^\infty S_\a(\ell)\, t^\ell, \hskip1cm \a=1,2. \label{gen}
\ee
Multiply (\ref{Q}) by $t^\ell$ and sum over $\ell$.  We obtain 
\bea
(1-at)G_1(t) -S_1(0) = \frac {A\,a^{-1}t}{1-a^{-1}t} + \frac {t+at^2} {N(1-a^2)(1-t^2)}.
\eea
This leads to
\bea
G_1(t) &=& \frac 1 {1-at} \Bigg[ S_1(0) +\frac {A\,a^{-1}t}{1-a^{-1}t} + 
\frac {t+at^2} {N(1-a^2)(1-t^2)} \Bigg] \nonumber \\
 &=&\frac 1{(1-a^2)(1-a^{2N})} \Bigg[\frac 1 {1-at}
  +\frac {a^{2N} }{1-a^{-1}t} \Bigg] \nonumber \\
&&+\, \frac 1 {2N(1-a)^2(1-t)} -\frac 1 {2N(1+a)^2(1+t)},\nonumber
\eea
from which one obtains
\bea
S_1(\ell) &=& \frac  {a^\ell +a^{2N-\ell}} {(1-a^2)(1-a^{2N})} 
+ \frac 1 {2N(1-a)^2} - \frac {(-1)^\ell}{2N(1+a)^2} \nonumber \\
&=& \frac  {a^\ell +a^{2N-\ell}} {(1-a^2)(1-a^{2N})} 
+\frac 1 {2N} \Bigg[\frac {4a} {(1-a^2)^2} +\frac {1-(-1)^\ell}{(1+a^2)^2}\Bigg].
\eea
It follows that using
 $I_1(\ell) = 2\,a\, S_1(\ell)$ we obtain (\ref{I1}) after setting $a=e^{-\l}$. QED
\hfill{$\bullet$}
 
\medskip
\noindent
2. Proof of (\ref{I2}):

\medskip
Again, we first evaluate $S_2(0)$ by carrying out the summation 
\be
{\cal {R}}e\, 
\frac 1 N \sum_{n=0}^{N-1}
  \frac 1 { 1- a\, e^{i2\p_n}}, \hskip1cm a<1
\ee
in two different ways.  First as in 
(\ref{I11}) we have 
\bea
{\cal {R}}e\, 
\frac 1 N \sum_{n=0}^{N-1}
  \frac 1 { 1- a\, e^{i2\p_n}}
  = \frac 1 2 \Big[1+ (1-a^2)  S_2(0)\Big], \label{I21}
\eea
where $S_2(\ell)$ is defined in (\ref{SL}).
 Secondly by expanding the summand we have 
\be
\frac 1 N \sum_{n=0}^{N-1}
  \frac 1 { 1- a\, e^{i2\p_n}}
=\frac 1 N \sum_{n=0}^{N-1} \sum_{\ell=0}^\infty a^\ell e^{i2\ell n\pi/N}
=\frac 1 {1-a^N} \label{I22}
\ee
where by carrying out the summation over $n$ 
for fixed $\ell$ all terms in (\ref{I12})
vanish except those with $\ell = m N, m=0,1,2,...$
 Equating  
(\ref{I22}) with (\ref{I21})  we obtain
\bea
  S_2(0)= \frac 1 {1-a^2} \Bigg(\frac {1+a^N} {1-a^N} \Bigg) \label{SN2}
\eea
and from (\ref{S10})
\bea
S_2(1) = \frac 1 {1-a^N}. \nonumber
\eea
We consider next the summation
\be
 {\cal {R}}e\, 
\frac {1} N \sum_{n=0}^{N-1}
  \frac {1-\big(a\, e^{i2 \p_n}\big)^\ell} { 1- a\, e^{i2 \p_n}} \hskip1cm a<1. \label{I20}
\ee
Evaluating the real part of the summand directly as in (\ref{I41}), we obtain
\bea
{\cal {R}}e\, 
\frac {1} N \sum_{n=0}^{N-1}
  \frac {1-\big(a\, e^{i2 \p_n}\big)^\ell} { 1- a\, e^{i2 \p_n}} =
  S_2(0) - a S_2(1) - a^\ell S_2(\ell) +a^{\ell+1} S_2(\ell-1).\label{I23}
 \eea
   Secondly, expanding the summand in (\ref{I20}) we obtain
\bea
\frac {1} N \sum_{n=0}^{N-1} 
  \frac {1-\big(a\, e^{i2 \p_n}\big)^\ell} { 1- a\, e^{i2 \p_n}} 
 &=& \frac {1} N \sum_{n=0}^{N-1}
  \sum_{m=0}^{\ell-1} a^m e^{i2\pi mn/N} \nonumber \\
&=& \frac 1 N \Bigg[ N + \sum_{m=1}^{\ell-1} \frac {1-e^{i2m\pi}}{1-e^{i2m\pi/N}}\Bigg] \nonumber \\
&=& 1 \hskip 2cm m<\ell \leq N . \label{I24}
\eea
 Equating (\ref{I24}) and (\ref{I23}) and making use of (\ref{SN2})
for $S_2(0)$, we obtain
\bea
S_2(\ell) -a S_2(\ell-1) = \frac {a^{N-\ell}}{1-a^N}  \label{rec1}
\eea
  
The recursion relation (\ref{rec1}) can be solved as in the above.
Define the generating function $G_2(t)$ by (\ref{gen}).
 We find
\bea
G_2(t) &=& 
\frac {1}{ 1-at}\Bigg[S_2(0) + \frac {a^{N-1}t}{(1-a^N) (1-a^{-1}t)} \Bigg] \nonumber \\
 &=& \frac 1 {(1-a^2)(1-a^{2N})} \Bigg[\frac 1 { 1-at} + \frac {a^N} { 1-a^{-1}t}\Bigg],
\eea
from which one reads off
\be
S_2(\ell) = \frac {a^\ell +a^{N-\ell}} {(1-a^2)(1-a^{2N})}.
\ee
Using the relation $I_2(\ell) = 2\, a\, S_2(\ell)$ with $ a=e^{-\l}$, we obtain (\ref{I2}). QED
\hfill{$\bullet$}

\section*{Acknowledgments}
I would like to thank D. H. Lee for discussions and
the hospitality at Berkeley
where this work was initiated.
 I am grateful to    W.-J. Tzeng for 
a critical reading of the
manuscript and helps in clarifying
 the M\"obius strip and Klein bottle analyses, and to W. T. Lu for assistance
in preparing the graphs. 
   Work is supported in part by NSF Grant DMR-9980440.

\newpage
{\center{\bf Figure Captions}}

\bigskip
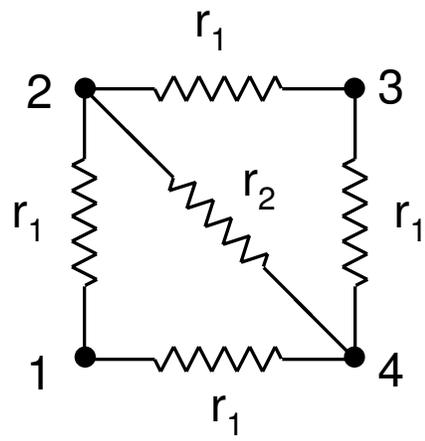
Fig. 1. A network of 4 nodes.

\bigskip
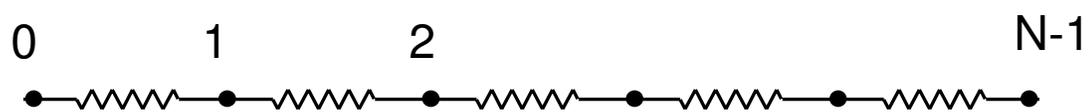
Fig. 2. A one-dimensional network of ${\cal N}$ nodes with  free ends.

\bigskip
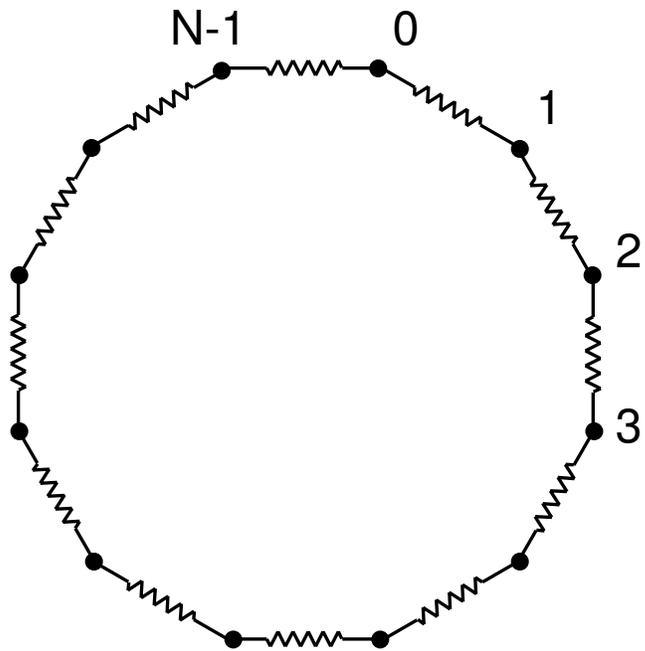
Fig. 3. A one-dimensional network of ${\cal N}$ nodes with  periodic boundary conditions

\bigskip
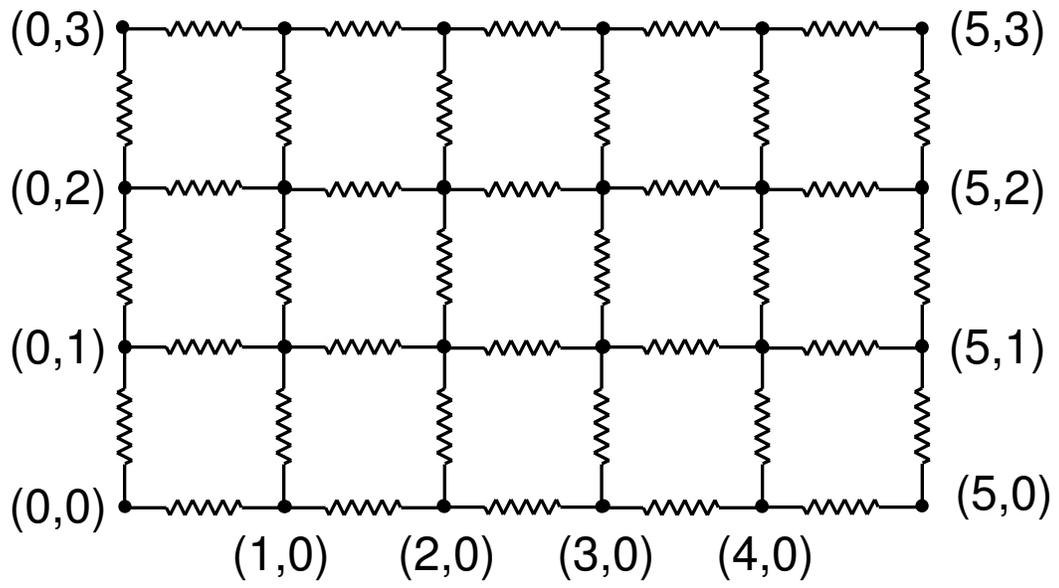
Fig. 4. A $5\times 4$ rectangular network.

\bigskip
Fig. 5 (a) The schematic plot of an M\"obius strip.
(b) The schematic plot of  a  Klein bottle.


\newpage
   \begin{thebibliography}{12}
\bibitem{cserti} J. Cserti, Application of the lattice Green's function for
calculating the resistance of an infinite network of resistors,
{\it Am. J. Phys.}
{\bf 68} 896-906  (2002).

\bibitem{ds} P. G. Doyle and J. L. Snell, Random walks and electric networks,
The Carus Mathematical Monograph, Series 22 (The Mathematical Association
of America, USA, 1984), pp. 83-149; Also in arXiv:math.PR/0001057.

\bibitem{lovasz} L. Lov\'asz, Random walks on graphs: A survey, in {\it Combinatorics,
Paul Erd\"os is Eighty}, Vol. 2, Eds. D. Mikl\'os, V. T. S\'os,
and T. Sz\'onyi (J\'anos Bolyai Mathematical Society, Budepest, 1996)
pp. 353-398. Also at http://research.microsoft.com/~lovasz/ as a survey paper.
  I am indebted to S. Redner for calling my attention to this
reference.
 


\bibitem{pol} B. van der Pol, The finite-difference analogy of the 
periodic wave equation and the potential equation, in {\it Probability
and Related Topics in Physical Sciences}, Lectures in Applied Mathematics,
Vol. 1, Ed. M. Kac (Interscience Publ. London, 1959) pp. 237-257.

\bibitem{redner} S. Redner, A Guide to First-passage Processes (Cambridge Press,
Cambridge, UK 2001).

 \bibitem{lg} 
S. Katsura, T. Morita, S. Inawashiro, T. Horiguchi, and Y. Abe,
Lattice Green's function: Introduction,
{\it J. Math. Phys.} {\bf 12} 892-895 (1971).



\bibitem{kirch}  G. Kirchhoff, \"Uber die Aufl\"osung der
Gleichungen, auf    welche man bei der Untersuchung der linearen
Verteilung galvanischer Str\"ome gef\"uhrt wird,
{\it Ann. Phys. und Chemie,} {\bf 72}
497-508 (1847).
 
\bibitem{note} A study of random walks using matrices can be found in \cite{lovasz}.
 
\bibitem{cserti1} J. Cserti, G. D\'avid, and Attila Pir\'oth,
Perturbation of infinite networks of resistors, {\it Am. J. Phys.}
{\bf 70} 153-159 (2002). Also in arXiv:cond-mat/0107362.


\bibitem{fara} See, for example, F. Harary, Graph theory (Addison-Wesley, Reading, MA 1969).

 

 
\bibitem{tw} W.-J. Tzeng and F. Y. Wu,
 Spanning Trees on Hypercubic Lattices and Non-orientable Surfaces,
 {\it Appl. Math. Lett.}  {\bf 13} (7) 19-25 (2000).


 
\bibitem{tzeng} I am indebted to W.-J. Tzeng for working out (\ref{mob}) and (\ref{kleineigen}).
  

\end{thebibliography}
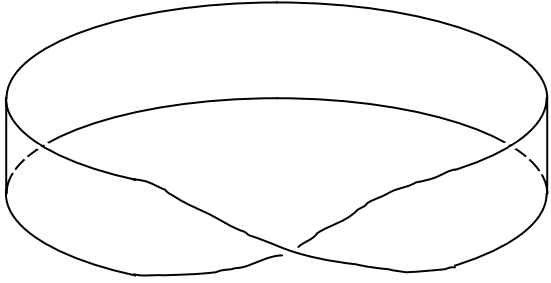
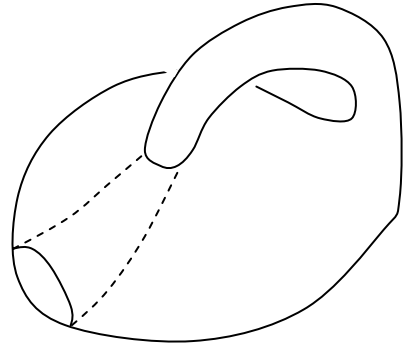
\end{document}